\documentclass[12pt]{article}

\usepackage[a4paper]{geometry}

\usepackage{latexsym}
\usepackage{amssymb}
\usepackage{euscript}
\usepackage{mathrsfs}
\usepackage{euscript}
\usepackage{xspace}
\usepackage{url}
\usepackage{tabularx}
\usepackage{booktabs}
\usepackage{graphicx}
\usepackage{xcolor}
\usepackage{enumerate}
\usepackage{soul}\sethlcolor{lightgray}

\usepackage{stackengine,scalerel}
\usepackage{color}

\usepackage[sort,longnamesfirst]{natbib}
\setcitestyle{authoryear,round,citesep={,},aysep={,},yysep={,},notesep={, }}

    \usepackage{tikz}
    \usetikzlibrary{patterns}
    \usepackage{intervalarrows}

\usepackage[pagewise]{lineno}
\input{ntheoremdefs}
\AtBeginDocument{\renewcommand{\iff}{\Leftrightarrow}}

\newcommand{\Not}[1]{\text{\textit{Not}}\lbrack #1 \rbrack}

\newcommand{\vide}{\varnothing}

\newcommand{\Max}{\operatorname{Max}}
\newcommand{\Min}{\operatorname{Min}}

\newcommand{\NN}{\Nat}

\newcommand{\Nat}{\mathbb{N}}

\newcommand{\wrt}{w.r.t.\ }
\newcommand{\resp}{resp.\ }
\newcommand{\ie}{i.e.\xspace}

\newcommand{\TRI}{ELECTRE TRI\xspace}
\newcommand{\TRIs}{ETRI\xspace}

\newcommand{\lTB}{ELECTRE TRI-B\xspace}
\newcommand{\lnTB}{ELECTRE TRI-nB\xspace}
\newcommand{\lnTBpc}{\lnTB, pseudo-conjunctive\xspace}
\newcommand{\lnTBpd}{\lnTB, pseudo-disjunctive\xspace}

\newcommand{\nTB}{\TRIs-nB\xspace}
\newcommand{\TB}{\TRIs-B\xspace}

\newcommand{\nTBpc}{\nTB-pc\xspace}
\newcommand{\nTBpd}{\nTB-pd\xspace}

\newcommand{\TBpc}{\mbox{\TB-pc}\xspace}
\newcommand{\TBpd}{\mbox{\TB-pd}\xspace}

\newcommand{\linear}{linear\xspace}

\newcommand{\D}{\ensuremath{(\D1)}}

\newcommand{\rS}{\mathrel{S}}

\newcommand{\rI}{\mathrel{I}}
\newcommand{\rP}{\mathrel{P}}
\newcommand{\rV}{\mathrel{V}}
\newcommand{\rU}{\mathrel{U}}
\newcommand{\rW}{\mathrel{W}}

\newcommand{\rSi}{\mathrel{S_{i}}}

\newcommand{\rIi}{\mathrel{I_{i}}}
\newcommand{\rPi}{\mathrel{P_{i}}}
\newcommand{\rVi}{\mathrel{V_{i}}}
\newcommand{\rUi}{\mathrel{U_{i}}}
\newcommand{\rWi}{\mathrel{W_{i}}}

\newcommand{\Set}{X}
\newcommand{\Seti}{\ensuremath{\Set_{i}}}
\newcommand{\N}{\ensuremath{N}}

\newcommand{\Aa}{\ensuremath{\mathcal{A}}}
\newcommand{\Uu}{\ensuremath{\mathcal{U}}}
\newcommand{\PART}{\ensuremath{\langle\Aa,\Uu\rangle}}

\newcommand{\As}{\Aa_{*}}
\newcommand{\Us}{\Uu^{*}}
\newcommand{\F}{\ensuremath{\mathcal{F}}}  

\newcommand{\Pro}{\mathcal{P}}

\newcommand{\relationnormale}{\succsim}
\newcommand{\arelationnormale}{\succ}
\newcommand{\srelationnormale}{\sim}

\newcommand{\qodom}{\relationnormale}
\newcommand{\aqodom}{\arelationnormale}

\newcommand{\es}{\srelationnormale}
\newcommand{\as}{\arelationnormale}

\newcommand{\relsi}{\relationnormale_{i}}

\newcommand{\asi}{\arelationnormale_{i}}

\makeatletter
\let\@fnsymbol\@alph
\makeatother

\title{\lnTBpd: axiomatic and combinatorial results%
\,\thanks{Authors are listed alphabetically. They have contributed equally.}
}
\author{Denis Bouyssou\,\thanks{Former Senior Researcher, CNRS, Paris, France,
e-mail: \protect\url{dbouyssou@gmail.com}. }
\and
Thierry Marchant\,\thanks{Ghent University, Department of Data Analysis,
H.\ Dunantlaan, 1, B-9000 Gent, Belgium,
e-mail: \protect\url{thierry.marchant@UGent.be}.}
\and
Marc Pirlot\,\thanks{Universit\'{e} de Mons, rue de Houdain 9, 7000 Mons, Belgium,
e-mail: \protect\url{marc.pirlot@umons.ac.be}.}}

\date{October 16, 2024}

\begin{document}
\maketitle


\begin{abstract}
\lnTB is a method designed to sort alternatives evaluated on several attributes into ordered categories. It is an extension of \lTB,  using several limiting profiles, instead of just one, to delimit each category. \lnTB comes in two flavours: pseudo-conjunctive and pseudo-disjunctive.
In a previous paper we have characterized the ordered partitions that can be obtained with \lnTBpc, using a simple axiom called linearity. The present paper is dedicated to the axiomatic analysis of \lnTBpd. It also provides some combinatorial results.

\smallskip

\end{abstract} 
\section{Introduction}\label{se:introduction}

\TRI (or \TRIs for short) is a family of methods for sorting alternatives evaluated on several attributes into ordered categories. The first  method in this family was \TB  \citep{YuWei92,RoyBouyssou93}. Then came several variants, that we do not detail.\footnote{For an overview of ELECTRE methods, we refer to
\citet[Ch.\ 5 \& 6]{RoyBouyssou93},
\citet{FigueiraELECTREJMCDA2013}, and \citet{FigueiraMousseauRoy2016}.}
Recently, \citet{FernandezEJOR2017} proposed a new variant (named \lnTB or \nTB for short)  that uses several limiting profiles instead of merely one as in the original \TB. Like  \TB, the new \nTB has two versions: pseudo-conjunctive (pc) and pseudo-disjunctive (pd). 

A simplified version of \TBpc received a detailed axiomatic analysis  in \cite{BouyssouMarchant2007:I,BouyssouMarchant2007:II,GMS02Control,GrecoMatarazzoSlowinski01Colorni}. Later, \cite{BouyssouMarchant2015} have shown that \TBpd is much more difficult to analyze than \TBpc, although their definitions may seem dual to each other at first sight.

\cite{BouyssouMarchantPirlot2023}---hereafter referred to as BMP23---have characterized the pseudo-conjunctive version of \nTB making auxiliary use of a simplified version thereof. This characterizations uses a single axiom---Linearity---that was first proposed by \cite{Goldstein91JMP}. \cite{BouyssouMarchantPirlot2022} have characterized the particular case of \nTB using at most 2 limiting profiles.

The present paper intends to axiomatically analyze \nTBpd or a simplified version thereof.
Our main findings are twofold. The first one is similar to that in \cite{BouyssouMarchant2015}: \nTBpd is much more difficult to analyze than \nTBpc, although their definitions may seem dual to each other. The second one is a characterization of a special case of \nTBpd, involving Linearity and a new condition, raising some interesting combinatorial questions about maximal antichains in direct products of chains.

\section{Framework and notation}
\label{se:Notation}

We use the framework of conjoint measurement \citep{KrantzLuceSuppesTversky71Vol1}. As in BMP23, we will restrict our attention to the case of two categories. This allows us to use a simple framework while not concealing any important difficulty.\footnote{\cite{BouyssouMarchant2007:II} have shown how to extend the axiomatic analysis to the case of more than two categories, in the case of \TB. Their technique applies mutatis mutandis to \nTB.} For the same reasons, we suppose throughout that the set of objects to be sorted is finite. 

The finite set of alternatives is $X = X_{1} \times \ldots \times X_{n}$, with $n \geq 2$. The set of attributes is $N= \{1, \ldots, n\}$. For $x,y \in X, i \in N$ and $J \subseteq N$, we use $X_{J}, X_{-J}, X_{i}, X_{-i}, (x_{J},y_{-J})$ and $(x_{i},y_{-i})$ as usual. Our primitives consist of a twofold partition $\PART$ of the set $X$, where $\Aa$ (resp. $\Uu$) contains the s\underline{A}tisfactory (resp. \underline{U}nsatisfactory) alternatives. 

An attribute $i$ is influential for $\PART$ if there exist $x_{i},y_{i} \in X_{i}$ and $a_{-i} \in X_{-i}$ such that $(x_{i},a_{-i}) \in \Aa$ and $(y_{i},a_{-i}) \in \Uu$. If an attribute is not influential, it does not play any role and can be suppressed. We therefore suppose without loss of generality that all attributes are influential.
\section{Axiomatic analysis of \nTBpc: a digest}
\label{se:digest}

In this section, we recall some definitions and results presented in BMP23.
All \TRIs methods start with a preference modelling step during which a preference relation is built for each attribute. This valued preference relation depends on a number of parameters that we do not detail here. 
In a second step, these $n$ valued preference relations are  aggregated into a single valued preference relation that is afterwards cut to define a crisp outranking relation $S$.
The assignment of alternatives to categories occurs in a third step. In order to save space, we do not present the exact definition of \nTBpc, but an idealization thereof: Model $E$. It mostly simplifies steps 1 and 2 and we will later see that this does not entail any loss of generality. See \citet{FernandezEJOR2017} for a complete description of \nTBpc and BMP23 for the relationship between Model E and \nTBpc.

\begin{definition}[Models $E, E^{c},E^{u}$]\label{def:model:M}
We say that a partition
$\PART$ has a representation in Model $E$ if:
\begin{itemize}
\item for all $i \in \N$, there is a semiorder
$\rSi$ on $X_{i}$ (with asymmetric part
$\rPi$ and symmetric part $\rIi$),
 
\item for all $i \in \N$, there is a strict semiorder $\rVi$
on $X_{i}$ that is included in $\rPi$ and is the asymmetric part of a semiorder
$\rUi$,

\item $(\rSi, \rUi)$ is a homogeneous nested chain of semiorders
and ${\rWi} $ is a weak order
that is compatible with both
$\rSi$ and $\rUi$,
\footnote{$\rW_i$ is the intersection of the weak orders  $\rS_{i}^{wo}$ and $\rU_{i}^{wo}$, respectively induced by $\rS_i$ and $\rU_i$. See Appendix A of the supplementary material of BMP23.}

\item there is a set of subsets of attributes  $\F \subseteq 2^{N}$ such that,
for all $I, J \in 2^{N}$, $[I \in \F$ and $I \subseteq J]$
$\Rightarrow J \in \F$,

\item there is a binary relation $\rS$ on $X$ (with
symmetric part $\rI$ and asymmetric part $\rP$) defined
by
\begin{equation*}
x \rS y \iff \left[S(x,y) \in \F \text{ and } V(y,x) = \vide\right],
\end{equation*}
where $S(x, y) = \{i \in \N : x_{i} \rSi y_{i}\}$ and
$V(x, y) = \{i \in \N : x_{i} \rVi y_{i}\}$,

\item there is a set $\Pro = \{p^{1}, \ldots , p^{k}\} \subseteq X$ of $k$
limiting profiles, 
such that for all $p, q \in \Pro$, $\Not{p \rP q}$,
\end{itemize}

such that

\begin{equation}\label{eq:M}
x \in \Aa  \iff
\begin{cases}
x \rS p    &\text{ for some } p \in \Pro \quad \text{and}\\
\Not{q \rP x}  &\text{ for all } q \in \Pro.
\end{cases}
\end{equation}

We then say that $\langle (\rSi, \rVi)_{i\in\N}, \F, \Pro\rangle$
is a representation of $\PART$ in Model $E$.
Model $E^{c}$ is the particular case of Model $E$,
in which there is a representation that
shows no discordance effects, \ie
in which all relations $\rVi$ are empty.
Model $E^{u}$ is the particular case of Model $E^{c}$, in which there is a representation
that requires unanimity, \ie such that $\F = \{N\}$.
\end{definition}
In this definition, $\rSi$ is the idealization of the preference relation on attribute $i$, $\rVi$ represents all pairs of levels on attribute $i$ for which a discordance could occur (step 1).\footnote{In Definition~\ref{def:model:M}, $\rSi$ and $\rVi$ are supposed to be semiorders. The reason of this assumption is that the notion of semiorder is related to the existence of thresholds, as they appear in the modelling of preference and veto in the classical ELECTRE methods.}   $S$ is the idealization of the outranking relation (step 2). The third step (the assignment of alternatives to categories) is described by \eqref{eq:M}.

\cite{Goldstein91JMP} has proposed a simple condition that may be satisfied by some partitions:
\begin{definition}[Linearity]\label{def:linearity}
The partition $\PART$ is linear on attribute $i$ if, for all $x_{i}, y_{i} \in X_{i}$ and all $a_{-i}, b_{-i} \in X_{-i},$
\begin{equation} \label{eq:lin}
\left.
\begin{array}{c}
(x_{i}, a_{-i}) \in \Aa \\
\text{and} \\
(y_{i}, b_{-i}) \in \Aa 
\end{array}
\right\}
\Rightarrow
\left\{
\begin{array}{c}
(y_{i}, a_{-i}) \in \Aa \\
\text{or} \\
(x_{i}, b_{-i}) \in \Aa.
\end{array}
\right.
\end{equation}
The partition $\PART$ is linear if it is linear on all attributes. If all partitions that can be represented in some Model $M$ are linear, we say that Model $M$ satisfies Linearity.
\end{definition}
Replacing $\Aa$ by $\Uu$ in \eqref{eq:lin} yields an equivalent definition of Linearity.
On each attribute $X_{i}$, we define the relation  $\relsi$ letting, for all $x_{i}, y_{i} \in X_{i}$, 
$$x_{i} \relsi y_{i} \text{ if } [ \text{for all } a_{-i} \in X_{-i}, (y_{i}, a_{-i}) \in \Aa \Rightarrow (x_{i}, a_{-i}) \in \Aa].$$
By construction, $\relsi$ is transitive and reflexive; it is complete if and only if the partition is linear on attribute $i$.
 The symmetric part of $\relsi$ is denoted by $\es_{i}$. It is not useful to keep in $X_{i}$ elements
that are equivalent \wrt the equivalence relation $\es_{i}$.
Indeed, if
$x_{i} \es y_{i}$ then $(x_{i}, a_{-i}) \in \Aa$ iff $(y_{i}, a_{-i}) \in \Aa$.
In order to simplify notation, we suppose throughout the paper that we are dealing with partitions
on $\prod_{i=1}^{n} \Seti$ for which all relations $\es_{i}$ are trivial\footnote{If $\es_{i}$ is not trivial, we can work without loss of generality with the quotient $X_{i}/\sim_{i}$.}. This non-restrictive convention implies that each relation $\succsim_{i}$ is antisymmetric.

Let $\succsim$ be the relation on $X$ defined by $x \succsim y$ iff $x_{i} \succsim_{i} y_{i}$ for all $i \in N$.  This relation is a partial order (reflexive, transitive and antisymmetric). Let $\As = \min(\Aa, \succsim)$ be the set of minimal elements in $\Aa$ for $\succsim$. By construction, for any $x \in \As$ and $y_{i} \prec_{i} x_{i}$, we have $(y_{i},x_{-i}) \in \Uu$.

We say Model $M$ is nested in---or is a special case of---Model $M'$ (denoted $M \subseteq M'$) if all partitions that can be represented in  $M$ can also be represented in $M'$. Models $M$ and $M'$ are equivalent (denoted $M \equiv M'$) if $M \subseteq M'$ and $M' \subseteq M$. 
We note $M \subsetneq M'$ if $M \subseteq M'$ and $M$ is not equivalent to $M'$.
By construction, we have $E^{u} \subseteq E^{c} \subseteq E$. 
The main results in BMP23 can now be summarized in the following theorem. 
\begin{theorem}\label{theo:summary}
\begin{enumerate}

\item  \nTBpc $\equiv E \equiv E^{c} \equiv E^{u}$. 

\item A partition $\PART$ has a representation in any of these models iff it is \linear.

\item  This representation can always be taken to be
$\langle ({\relsi}, {\rVi} = \vide)_{i \in \N}, \F = \{\N\}, \Pro = \As \rangle$, that is a representation in Model  $E^{u}$.

\end{enumerate}
\end{theorem}
We like to stress  point 1: although model $E$ and the nested models $E^{c}$ and $E^{u}$ seem to be  simplifications of \nTBpc, they are not: all four models are fully equivalent.

\section{\nTBpd: definition and difficulties}\label{se:results}

The pseudo-disjunctive version of \nTB consists of  three steps. The first and the second one are identical to steps 1 and 2 in  \nTBpc. The only difference is the third step: the assignment of alternatives to categories.
With  \nTBpc, an alternative $x$ is assigned to $\Aa$ iff it is weakly preferred (in terms of $\rS$) to a limiting profile and no limiting profile is strictly preferred to $x$ (in terms of $\rP$), as in \eqref{eq:M}. With \nTBpd,  an alternative $x$ is assigned to $\Uu$ iff (i) there is a limiting profile strictly preferred (in terms of $\rP$) to $x$ and (ii) $x$ is not strictly preferred to any limiting profile.
As in  Section~\ref{se:digest}, in order to save space, we do not present the exact definition of \nTBpd, but an idealization thereof: Model $F$. 
We define Model $F$ that is to \nTBpd
what Model $E$ is to \nTBpc.

\begin{definition}\label{def:model:MM}
Model $F$ is defined exactly as Model $E$,
except that we now replace \eqref{eq:M}  by:
\begin{equation}\label{eq:MM}
x \in \Uu  \iff
\begin{cases}
p \rP x    &\text{ for some } p \in \Pro \quad \text{and}\\
\Not{x \rP q}  &\text{ for all } q \in \Pro.
\end{cases}
\end{equation}
The definition of Models
$F^{c}$ and $F^{u}$ parallels that of $E^{c}$ and $E^{u}$.
\end{definition}
All pseudo-disjunctive models mentioned so far satisfy linearity.
\begin{lemma}\label{lem:linearity}
If $\PART$ has a representation in Model $F$, then it is \linear. The same holds for \nTBpd,  $F^{u}$ and $F^{c}$.
\end{lemma}
\begin{proof}
Consider first Model $F$. 
Suppose that we have
$(x_{i}, a_{-i}) \in \Uu$ and
$(y_{i}, b_{-i}) \in \Uu$.
We have either $x_{i} \rWi y_{i}$ or $y_{i} \rWi x_{i}$ since $\rWi$ is a weak order.
Suppose wlog that $y_{i} \rWi x_{i}$.
Because $(y_{i}, b_{-i}) \in \Uu$, we know that
$p \rP (y_{i}, b_{-i})$, for some  $p \in \Pro$,
and
$\Not{ (y_{i}, b_{-i}) \rP q}$  for all $q \in \Pro$. Using Lemma~3 in BMP23, we obtain
$p \rP (x_{i}, b_{-i})$ and
$\Not{ (x_{i}, b_{-i}) \rP q}$  for all $q \in \Pro$.\footnote{ Lemma 3 in BMP23 is established under the hypothesis that the partition $[\Aa,\Uu]$ is representable in Model E. Since the proof only uses the properties of relation $\rS$, which are common to Models E and F, the result also holds  for partitions representable in Model F. }
Hence, $(x_{i}, b_{-i}) \in \Uu$ and linearity holds for Model $F$. By construction, $F^{u} \subseteq F^{c} \subseteq F$ and linearity thus also holds for these models.

Since we did not  formally define  \nTBpd, we cannot provide the proof that linearity holds for partitions generated by \nTBpd. For the interested reader, this proof closely follows  that of Corollary~1 in BMP23. 
\end{proof}
Hence, combining Lemma~\ref{lem:linearity}
with Theorem~\ref{theo:summary}, we obtain the next proposition.
\begin{proposition}
\label{prop:EF}
$F^{u} \subseteq F^{c} \subseteq F \subseteq E$ and \nTBpd $\subseteq E \equiv$ \nTBpc.
\end{proposition}

At this stage, given the apparent duality between the definitions of the pseudo-conjunctive and pseudo-disjunctive models, we can suspect that $F^{u} \equiv F^{c} \equiv F \equiv$ \nTBpd $\equiv E$, but the next result shows that it does not hold. 

\begin{proposition}
\label{prop:FuNeqFc}
$F^{u} \subsetneq F^{c}$ and $F \subsetneq E$.
\end{proposition}

\begin{proof}
\subsubsection*{Part 1: $F^{u} \subsetneq F^{c}$}
Let $\N = \{1, 2, 3\}$ and $\Seti = \{0, 1\}$ for all $i \in \N$, so that $\Set$
has $2^{3}=8$ elements.
Consider the partition $\PART$ such that
$\Aa = \{111, 101, 011\}$
and
$\Uu =\{110, 100, 010, 001, 000\}$,
abusing notation in an obvious way.
It is simple to check that all attributes are influential for $\PART$
and that, for all $i \in \N$, we have
$1_{i} \asi 0_{i}$.
Notice that we have
$\As = \Min(\qodom, \Aa)$ $=$ $\{101, 011\}$ and
$\Us = \Max(\qodom,\Uu) =$ $\{110, 010, 001\}$.

Let us show that this partition \emph{cannot} be obtained with Model
$F^{u}$.
Observe first that, here, since all attributes are influential and can only take two values,
we must have that ${\rS_{i}} = {\relsi}$, for all $i \in \N$.

Since $110 \in \Uu$, there must be $p \in \Pro$ such that
$p \rP 110$. Since we are looking for a representation in Model
$F^{u}$ and we know that ${\rS_{i}} = {\relsi}$, for all $i \in \N$, we must find
a profile $p \in \Pro$ such that $p \aqodom 110$.
The only candidate is $111$. But taking
$\Pro = \{111\}$ together with $\F = \{\N\}$ does
not lead to the desired partition. Indeed, we have
$111 \aqodom 101$, so that $101$ should be in $\Uu$.

This partition can be obtained with Model $F^{c}$, taking
${\rS_{i}} = {\relsi}$, for all $i \in \N$,
$\Pro =\{111\}$ and $\F =\{\{1,3\}, \{2,3\}\}$.

\subsubsection*{Part 2: $F \subsetneq E$}
Let $n = 4$ and $\Set_{1} = \Set_{2} = \Set_{3} = \{2, 1, 0\}$ and
$\Set_{4} = \{0, 1\}$, so that $\Set$ has $54$ elements.
Consider the partition $\PART$ such that
$\Aa = \{2221$, $2211$, $2121$, $1221$, $2111$, $1211$, $1121$,
$1111$, $2220\}$.
Notice that   $\As = \{1111, 2220\}$.
It is easy to check that all attributes are influential for $\PART$
and that, for all $i \in \{1,2,3\}$, we have
$2_{i} \asi 1_{i} \asi 0_{i}$, while $1_{4} \as_{4} 0_{4}$.
Hence, the partition is linear and, by Theorem~\ref{theo:summary}, it can be represented in Model $E$.

In order to show that this partition cannot be obtained in
Model $F$, we have to examine, all cases
of indifference thresholds (associated with the strict semiorders $\rSi$), combined with all cases of veto thresholds (associated with the strict semiorders $\rVi$), and
combined with
all choices for $\F$.

Notice that if an attribute in $i \in \{1,2,3\}$ has thresholds (i.e. $\rSi$ is not a weak order),
this means that $2_{i} \rI_{i} 1_{i}$ and $1_{i} \rI_{i} 0_{i}$. But
veto effects can only occur among the elements that are strictly preferred.
Hence, in this case, the only possibility is to take
$2_{i} \rV_{i} 0_{i}$.

If $\{1,2,4\} \in \F$, then, without veto,
$2201 \in \Uu$ outranks all elements in $\Aa$, a contradiction.
This will remain true unless, there is a veto effect on attribute $3$.

If $2_{3} \rV_{3} 0_{3}$,
the only elements in
$\Aa$ that are not strictly beaten by anther element in $\Aa$ are
$2220$, $1111$, and $1121$.
It is easy to check that taking all of them or any subset of them
as the set of profiles does not lead to the desired partition
(consider $2201 \in \Uu$).
If, furthermore, $2_{3} \rV_{3} 1_{3}$,
the only elements in
$\Aa$ that are not strictly beaten by anther element in $\Aa$ are
$2220$ and $1111$.
It is easy to check that taking all of them or any subset of them
as the set of profiles does not lead to the desired partition
(consider $2201 \in \Uu$).

The analysis of the cases $\{1,3,4\} \in \F$ and
$\{2,3,4\} \in \F$ is entirely similar.

\medskip

Suppose now that $\F = \{\{1,2,3\}, \N\}$.
Suppose that only attribute $1$ has thresholds.
Without veto, it is easy to check that $1220 \in \Uu$ outranks all elements in $\Aa$.
This remains true, whatever the choice of veto thresholds on attributes
$2$ and $3$. This also remains true if $1_{4} \rV_{4} 0_{4}$.
But veto effects on attribute $1$ are immaterial since $1_{1}$ is indifferent
to both $2_{1}$ and $0_{1}$.

The situation is entirely similar if $2$ (\resp $3$) is the only attribute to have
thresholds.

Suppose that only attributes
$1$ and $2$
have thresholds.
Without veto,
it is easy to check that
$1120 \in \Uu$
outranks all elements in $\Aa$.
This remains true, whatever the choice of veto thresholds on attributes
$1$ and $2$ since $1_{1}$ (\resp $1_{2}$) is indifferent to $2_{1}$ and $0_{1}$
(\resp $2_{2}$ and $0_{2}$).
This also remains true if  $1_{4} \rV_{4} 0_{4}$.
Clearly, the veto threshold on attribute $3$ is immaterial.

The analysis of the cases in which $1$ and $3$ or $2$ and $3$ have thresholds is entirely similar.

\medskip

It remains to tackle the case
$\F = \{\N\}$.

Suppose that only attribute $1$ has thresholds.
Without veto,
there are only $3$ elements in $\Aa$ that are not strictly beaten by another element
in $\Aa$: $2220$, $1111$ and $2111$.
It is easy to check that taking all of them or any subset of them
as the set of profiles does not lead to the desired partition.
It is simple to check that whatever the choice of veto we make on attributes
$2$, $3$ and $4$, the situation remains the same.

There is only one possibility to put a veto on attribute $1$, \ie
$2_{1} \rV_{1} 0_{1}$. In this case
there are only $2$ elements in $\Aa$ that are not strictly beaten by another element
in $\Aa$: $2220$ and $1111$.
In any case, it is impossible to recover the desired partition.

The situation is entirely symmetric
in the case only attribute $2$ or only attribute $3$ has thresholds.

Suppose that both attributes $1$ and $2$ have thresholds.
Without veto,
there are only $5$ elements in $\Aa$ that are not strictly beaten by another element
in $\Aa$: $2220$, $1111$, $2211$, $2111$, and $1211$.
It is easy to check that taking all of them or any subset of them
as the set of profiles does not lead to the desired partition.
It is simple to check that whatever the choice of veto we make on attributes
$3$ and $4$, the situation remains the same.
There is only one possibility to put a veto on attribute $1$ (\resp $2$), \ie
$2_{1} \rV_{1} 0_{1}$ (\resp $2_{2} \rV_{2} 0_{2}$).
It is simple  to check that any of the three possible choices for the veto on these attributes
does not alter the situation.

The situation is entirely symmetric
in the case only attributes $1$ and $3$ or only attributes $2$ and $3$ have thresholds.

Finally, if all attributes have thresholds,
there is only one element in $\Aa$ that is not strictly beaten by another element
in $\Aa$: $2220$.
It is easy to check that taking this element to be the unique profile,
does not lead to the desired partition.
Now, the choice of veto thresholds (they must be of the type
$2_{i} \rV_{i} 0_{i}$)
on attributes $1$, $2$, and $3$
is immaterial.
But it is also simple to check that adding a veto on attribute $4$
does not change the situation.
\end{proof}

Given Propositions~\ref{prop:EF} and~\ref{prop:FuNeqFc}, it would be highly desirable to know whether  \nTBpd $\equiv E$ or \nTBpd $\equiv F$. Unfortunately, we are presently unable to prove or disprove these equivalences.
This shows that the relations between the  pseudo-disjunctive models are more complex than between  the corresponding pseudo-conjunctive models.

\section{Two characterizations}
\label{se:characterizations}

In view of the above-metioned difficulties, we devote this section to two simpler problems: (1) the characterization of Models $F^{c}, F, E$ and \nTBpc when all attributes are binary and (2) the characterization of a special case of Model $F^{u}$.

\subsection{The case of binary attributes}

Suppose the partition $\PART$ is linear on attribute $i$. We say  attribute $i$ is binary if the weak order $\succsim_{i}$ has exactly two equivalence classes. Such attributes are common in many applications. The case in which all attributes are binary corresponds to the well-developed theory of monotone Boolean functions \citep[see][]{CramaHammer2011}.

\begin{proposition}\label{prop:MMu}
$F^{c} \equiv F \equiv E \equiv$ \nTBpc whenever all attributes are binary.
\end{proposition}

\begin{proof}
When all attributes are binary, each $X_i$ contains only two elements that we can denote by $1_i$ and $0_i$ with $1_i \succ_i 0_i$ .
Each  element in $X$ corresponds to a unique coalition $C(x)= \{i \in N : x_i = 1_i\} \subseteq 2^{\N}$. Hence, all linear partitions
have a representation in $F^{c}$ with ${\rS_{i}} = {\relsi}$, for all $i \in \N$,
$\Pro =\{111\}$ and $\F =\{C(x):x \in \Aa \}$.
\end{proof}

Since the proof uses a set  $\Pro$ containing only one limiting profile, the reader may have the impression that Proposition~\ref{prop:MMu} only applies to \TB and not to \nTB. What the result actually says is that any partition generated by a model $F^{c}, F, E$ or \nTBpc (irrespective of the number of limiting profiles) can be represented in the other three models. The proof further shows that the representation can be chosen so that $\Pro$ is a singleton.

\subsection{A  special case of Model $F^{u}$}

In order to reduce the complexity of the models, let us assume that the data are of good quality---in the sense of \citet[Section 8.2]{Roy96BookTranslated}---meaning that there is no imprecision, uncertainty, or inaccurate determination. In that case, there is no need to use preference or indifference thresholds and the relation ${\rSi}$ is a weak order. Since $\succsim_{i}$ is also a weak order and we cannot have $x_{i} \succ_{i} y_{i}$ while $y_{i} \rSi x_{i}$, it must be the case that ${\rSi}$ is a refinement of $\succsim_{i}$ (i.e.\ ${\rSi} \subseteq {\succsim_{i}}$). But since we have assumed that  the relation $\sim_{i}$ is trivial, the equality ${\rSi}= {\succsim_{i}}$ must hold.

So, in this section, we   restrict our attention to partitions having a representation in Model $F^{u}$  such that ${\rSi}={\succsim_{i}}$ is a weak order for all $i \in \N$.
Model  $F^{u}$
together with this additional constraint will be denoted by
 $F^{\underline{u}}$.
 In such a model, $S= \ \succsim$, $P= \ \succ$ and Condition~\ref{eq:MM} reduces to $x \in \Uu$ iff $p \succ x$, for some $p \in \Pro$. Indeed, we may not have $p \succ x \succ q$ for $q \in \Pro$, otherwise $p \succ q$, a contradiction.
Notice that $F^{\underline{u}} \subseteq F^{u} \subsetneq F^{c} \subseteq F \subsetneq E$.

By construction, the set $\As = \Min(\qodom, \Aa)$
is an antichain in the poset $(X, \qodom)$, remembering our convention
that each relation $\sim_{i}$ is trivial.
Observe that in the first part of the proof of Proposition~\ref{prop:FuNeqFc},
the antichain $\As = \{101, 011\}$ is not a maximal antichain,
\ie it is strictly included in the antichain
 $\{110, 101, 011\}$.
As shown below, a characteristic feature of partitions that can be represented in
Model $F^{\underline{u}}$ is that $\As$ is a maximal antichain in the poset $(X, \qodom)$.

\begin{theorem}\label{th:model:Fu}
Let $\Set = \prod_{i=1}^{n} \Seti$ be a finite set and $\PART$
be a twofold linear partition of $\Set$.
The partition $\PART$ has a representation in Model $F^{\underline{u}}$
iff the antichain $\As$, in the
poset $(X, \qodom)$, is maximal.
\end{theorem}

\begin{proof}
Necessity. Suppose that $\As$ is not a maximal antichain.
Hence there is $x \in X$
such that $x$ is incomparable, using $\qodom$, \wrt all elements
in $\As$. In view of the definition of $\As$, it is impossible that
$x \in \Aa$ (since this would imply that $x \qodom z$, for some $z \in \As$).
Hence, we must have $x \in \Uu$, so that
there must be a profile $p \in \Pro$ such that $p \aqodom x$.
This profile must be in $\Aa$. But, by hypothesis, this profile cannot belong to $\As$.
Hence, by construction, we know that $p \aqodom y$, for some
$y \in \As \subseteq \Aa$, which implies  $y \in \Uu$, a contradiction.

Sufficiency. Since $\PART$ is linear, we know that it has a representation in Model
$E^{u}$ using the representation
$\langle ({\relsi}, {\rVi} = \vide)_{i \in \N}, \F = \{\N\}, \Pro = \As \rangle$.
Since $\As$ is a maximal antichain,
it is easy to see that this representation is also a representation
in Model $F^{\underline{u}}$.
Indeed, by construction, it is impossible that
$x \in \Uu$ is incomparable, using $\qodom$, to all $p \in \Pro = \As$.
Let $q \in \Pro$ be such that $x$ and $q$ are comparable using $\qodom$.
It is impossible that $x \qodom q$ since this would imply that
$x \in \Aa$, in view of the definition
of $\As = \Pro$.
Hence, we must have that $q \aqodom x$.
\end{proof}

The next result shows that $F^{\underline{u}} \subsetneq F^{u}$, thereby showing that
the hypothesis that the representation is such that
${\rSi} = {\relsi}$, for all $i \in \N$, is \emph{not} innocuous.

\begin{proposition}\label{prop:FusFu}
 $F^{\underline{u}} \subsetneq F^{u}$.
\end{proposition}

\begin{proof}
Let $\N = \{1, 2, 3, 4\}$
and $\Seti = \{0, 1, 2\}$ for all $i \in \N$, so that $\Set$
has $3^{4} = 81$ elements.
Consider the partition $\PART$ such that
$\Aa = \{
 2222$,
$2221$,
$2220$,
$2212$,
$2211$,
$2210$,
$2202$,
$2201$,
$2200$,
$2122$,
$2121$,
$2120$,
$2112$,
$2111$,
$2110$,
$2102$,
$2101$,
$2022$,
$2021$,
$2020$,
$2012$,
$2011$,
$2010$,
$2002$,
$2001$,
$1222$,
$1221$,
$1220$,
$1212$,
$1211$,
$1210$,
$1202$,
$1201$,
$1200$,
$1122$,
$1121$,
$1120$,
$1112$,
$1111$,
$1110$,
$1102$,
$1101$,
$1022$,
$0222$,
$0221$,
$0220$,
$0212$,
$0211$,
$0210$,
$0202$,
$0201$,
$0122$,
$0121$,
$0120$,
$0112$,
$0111$,
$0110$,
$0102$,
$0101$,
$0022\}$.
The set $\Aa$ has $60$ elements. It is easy to check that
we have
$\As =\{
2010$,
$2001$,
$1200$,
$0110$,
$0101$,
$0022\}$.

We have
$2012 \in \Aa$, $1012 \in \Uu$,
$1200 \in \Aa$, $0200 \in \Uu$, so that
$2_{1} \as_{1} 1_{1} \as_{1} 0_{1}$.
Similarly, we have:
$2200 \in \Aa$, $2100 \in \Uu$,
$1101 \in \Aa$, $1001 \in \Uu$, so that
$2_{2} \as_{2} 1_{2} \as_{2} 0_{2}$.
We also have:
$0022 \in \Aa$, $0012 \in \Uu$,
$2110 \in \Aa$, $2100 \in \Uu$, so that
$2_{3} \as_{3} 1_{3} \as_{3} 0_{3}$.
Finally, we have:
$0022 \in \Aa$, $0021 \in \Uu$,
$2101 \in \Aa$, $2100 \in \Uu$, so that
$2_{4} \as_{4} 1_{4} \as_{4} 0_{4}$
(notice that the role of attributes $3$ and $4$ is entirely symmetric, in this example).


Hence, using Theorem~\ref{th:model:Fu}, this partition
\emph{cannot} be represented in Model $F^{\underline{u}}$. Indeed, the antichain
$\As$ is not maximal: the element $2100$ is incomparable, using $\qodom$, to all elements
in $\As$.

Yet it is cumbersome but easy to check that this partition can be obtained in Model $F^{u}$,
taking $\Pro = \{2200, 0022\}$, $\F =\{\N\}$,
 ${\rSi} = {\relsi}$, for $i = 2, 3, 4$,
 and
$2_{1} \rP_{1} 0_{1}$, $2_{1} \rI_{1} 1_{1}$, and
$1_{1} \rI_{1} 0_{1}$.
\end{proof}

Let us define $E^{\underline{u}}$ in the same way as $F^{\underline{u}}$. By Theorem~\ref{theo:summary}, $E^{\underline{u}}$ is equivalent to  $E^{u}$.
Summarizing Proposition~\ref{prop:FusFu} and previous results, 
we have that $F^{\underline{u}} \subsetneq F^{u} \subsetneq F^{c} \subseteq F \subsetneq E \equiv E^{c} \equiv E^{u} \equiv E^{\underline{u}}$. This long chain of inclusions and equivalences illustrates the strong asymmetry between the families of pseudo-conjunctive and pseudo-disjunctive models. In order to explore the gap between both families, we devote the rest of the paper to comparing the numbers of partitions that can be represented in models $F^{\underline{u}}$ and $E^{\underline{u}}$ (or any of the pseudo-conjunctive models discussed in this paper). This will help us quantify how restrictive  $F^{\underline{u}}$ is compared to $E^{\underline{u}}$.

\section{Counting maximal antichains}
\label{se:counting}

The number of partitions that can be represented in model $F^{\underline{u}}$ (resp.\ model $E^{\underline{u}}$) is  the number of maximal antichains (resp.\  antichains) in the poset $(X, \qodom)$.
This poset  can be seen as a direct product of $n$ chains, where $n$ is the number of attributes and the $i$th chain ($i \in \{1, \dots,n\}$) is the set $[m_{i}]=\{1, \ldots, m_{i}\}$ ordered by $\geq$ (the natural order on the integer interval $[m_{i}]$), with $m_{i}$ being the number of equivalence classes of the weak order $\succsim_{i}$. Notice that antichains in the direct product of $n$ chains also plays an important role in the analysis of multichoice cooperative games, as shown by \cite{Grabisch2016}. More generally, the importance of studying discrete mathematics structures in decision theory was powerfully stressed in \cite{GrabischBook2016}.

The number of antichains (maximal antichains) in $[m_{1}] \times \ldots \times [m_{n}]$ will be denoted by $d_{E}(m_{1}, \ldots, m_{n})$ (resp. $d_{F}(m_{1}, \ldots, m_{n})$).
When $m_{1}= \ldots = m_{n}=m$, the numbers $d_{E}(m_{1}, \ldots, m_{n})$ and $d_{F}(m_{1}, \ldots, m_{n})$ are respectively denoted by $D_{E}(m,n)$ and $D_{F}(m,n)$.
We first tackle two special cases ($n=2$ and $m=2$) and then the general case, for which we have  few results.

\subsection{The case $n=2$}

Let  $\NN$ denote the set of positive integers. The next result, due to \cite{Covington2004}, presents a recurrence relation for $d_{F}(m_{1},m_{2})$.

\begin{theorem}
\label{theo.df.m1.m2}
For all $m_{1},m_{2} \in \NN$, $d_{F}(m_{1},m_{2})$ is equal to
\begin{equation}
\label{eq.dF.m1.m2}
d_{F}(m_{1}-1,m_{2}-1) + \sum_{i=0}^{m_{1}-2} d_{F}(i,m_{2}-1) + \sum_{i=0}^{m_{2}-2} d_{F}(m_{1}-1,i).
\end{equation}
\end{theorem}
A detailed proof of this result can be found in \cite{BouyssouMarchantPirlot2024}.
For $d_{E}(m_{1},m_{2})$, the following result easily follows  from \cite{BermanKohler1976}.

\begin{corollary}
\label{prop.DE}
For all $m_{1}, m_{2} \in \NN$, we have 
$$ d_{E}(m_{1},m_{2}) =    \binom{m_{1}+m_{2}}{m_{1}}.$$
\end{corollary}

\begin{proof}
According to \cite{BermanKohler1976}, the number of antichains in $[m_{1}] \times [m_{2}] \times [m_{3}]$ is equal to 
\begin{equation}
\label{eq.BermanKohler}
\prod_{i=0}^{m_{3}-1} \frac{  \binom{m_{1}+m_{2}+i}{m_{1}} }{ \binom{m_{1}+i}{m_{1}}  }.
\end{equation}
Setting  $m_{3}=1$ in this expression yields the desired result.
\end{proof}

For illustration purpose, we computed some numerical results under the constraint that $m_{1}=m_{2}$ (to save space).
Some terms of the sequences $D_{E}(m,2)$ and $D_{F}(m,2)$ can be found in Table~\ref{table.examples.size.m2}, with
the  corresponding  ratios  $D_{F}(m,2)/D_{E}(m,2)$. For small values of $m$, the difference of expressivity between models $F^{\underline{u}}$ and $E^{\underline{u}}$ is not very large, but it grows for large values of $m$, since the ratio seems to converge to 0.

\begin{table}[hbp]
    \centering
    \begin{tabular}{rrrl}
        \hline
	$m$ & $D_{F}(m,2)$ & $D_{E}(m,2)$ &  $D_{F}(m,2)/D_{E}(m,2)$ \\
        \hline
        1	&	1 & 2 	& 0.5  \\
        2	&	3 & 	6 & 0.5  \\
        3	&	9 & 	20 & 0.45 \\
        4	&	27 & 	70 &  0.385714286\\
        5	&	83 & 	252 &  0.329365079 \\
        6	&	259 & 	924 &  0.28030303 \\
        7	&	817 & 	3432 & 0.238053613  \\
        8	&	2599 & 	12870 &  0.201942502 \\
        9	&	8323 & 	48620 &  0.171184698 \\
        10	&	26797 & 	184756 &  0.145039945 \\
        11	&	86659 & 	705432 & 0.122845292 \\
        12	&	281287 & 	2704156 &  0.104020256 \\
        13	&	915907 & 	10400600 & 0.088062900 \\
        14	&	2990383 & 	40116600 & 0.074542284  \\
        15	&	9786369 & 	155117520 &  0.06309003 \\
        100	&	3.76527E+51 & 	9.05485E+58 &  4.15829E-08 \\
        \hline
    \end{tabular}
    \caption{Number $D_{F}(m,2)$ of maximal antichains, number $D_{E}(m,2)$ of antichains  and ratio of these numbers in $[m]^{2}$ for $m \in [15]$ and $m=100$. Values of $D_{F}(m,2)$ are computed by means of \eqref{eq.dF.m1.m2}. }
    \label{table.examples.size.m2}
\end{table}

$D_{F}(m,2)$ and $D_{E}(m,2)$ are respectively sequences A171155 and A000984  in the On-line Encyclopedia of Integer Sequences \cite{OEIS2023}. A  recurrence relation is mentioned by Alois P. Heinz (without proof)  for $D_{F}(m,2)$ in \cite{OEIS2023}: $D_{F}(m,2)$ is equal to
$$
 \frac{(4m-3)D_{F}(m-1,2)-(2m-5)D_{F}(m-2,2)+D_{F}(m-3,2)-(m-3)D_{F}(m-4,2)}{m}.
$$

Some other results (old and new) about the case $n=2$ are presented in \cite{BouyssouMarchantPirlot2024}. Therein, in addition to enumeration results, correspondences (bijections) between (maximal) antichains in products of chains and other mathematical structures are established.

\subsection{The case $m=2$}

$D_{F}(2,n)$ is sequence A326358 in  \cite{OEIS2023}. No expression seems to be known for this sequence and the highest known value corresponds to $n=7$. Some terms can be found in Table~\ref{table.examples.size.2n}. 

$D_{E}(2,n)$ corresponds to the Dedekind numbers (sequence A000372 in \cite{OEIS2023}), for which no expression is known. The highest known value corresponds to $n=9$.
Some terms can be found in Table~\ref{table.examples.size.2n} with the
 corresponding ratios $D_{F}(2,n)/D_{E}(2,n)$. Here again, for small values of $n$, the difference of expressivity between models $F^{\underline{u}}$ and $E^{\underline{u}}$ is not very large, but for large values of $n$, the ratio seems to converge to 0.

\begin{table}[hbp]
    \centering
    \begin{tabular}{rrrl}
        \hline
	$n$ & $D_{F}(2,n)$ & $D_{E}(2,n)$ &  $D_{F}(2,n)/D_{E}(2,n)$ \\
        \hline
        1	&	2 & 3 	& 0.6666667  \\
        2	&	3 & 	6 & 0.5  \\
        3	&	7 & 	20 & 0.35 \\
        4	&	29 & 	168 &  0.172619 \\
        5	&	376 & 	7581 &  0.04959768 \\
        6	&	31746 & 	7828354 &  0.004055259 \\
        7	&	123805914 & 	2414682040998 & 0.00005127214  \\
        \hline
    \end{tabular}
    \caption{Number $D_{F}(2,n)$ of maximal antichains, number $D_{E}(2,n)$ of antichains  and ratio of these numbers in $[2]^{n}$ for $n \in [7]$.}
    \label{table.examples.size.2n}
\end{table}

\subsection{The general case}

In the general case, analytic expressions for $D_{F}(m,n)$ and $D_{E}(m,n)$ are difficult to obtain and we therefore only provide a lower bound for $D_{F}(m,n)$ and some numerical results.

\subsubsection{A lower bound for $D_{F}(m,n)$}


\begin{proposition}\label{prop:boundDF}
The number of maximal antichains in $[m]^{n}$ is at least the number of antichains of $[m]^{n-1}$, that is
$D_F(m,n) \geq D_E(m,n-1). $
\end{proposition}

\begin{proof}
The set $\{x \in [m]^n : x_i = m\}$ is the set of elements $x \in X$ having their $i$th coordinate equal to $m$. We shall prove that any antichain, not necessarily maximal, in $\{x \in [m]^n : x_i = m\}$ can be extended into a maximal antichain of $X$, which has no other element with its $i$th coordinate $x_i$ equal to $m$.
This will establish Proposition~\ref{prop:boundDF} since any antichain of $X_{-i}$ is in one-to-one correspondence with an antichain of $\{x \in [m]^n : x_i = m\}$.

We take wlog $i=1$. If the antichain in $\{x \in [m]^n : x_1 = m\}$ is maximal, the result is obvious.
Otherwise, let $A$ be any non-maximal antichain in $\{x \in [m]^n : x_1 = m\}$. Since $A$ is not maximal in $\{x \in [m]^n : x_1 = m\}$, there is at least one  element $x = (m,x_2, \ldots, x_n)$ that is incomparable to all elements in $A$. Let $x' = (m-1, x_2, \ldots, x_n)$. We have that $x \succ x'$ and $x'$ is incomparable to any element in $A$. Indeed, for no $y \in A$, we have $x' \succsim y$ (otherwise $x \succsim y$ would hold too) and, for no $y \in A$, we have $y \succsim x'$ (otherwise $y \succsim x$ would also hold). 
Consider the set of all elements in $\{x \in [m]^n : x_1 = m\}$ that are incomparable to all elements in $A$. Select the minimal elements from this set. Change the first coordinate of each minimal element $x$ into $x_1 = m -1$, yielding an element $x'$.  Let $A'$ be the set obtained by adding all such elements $x'$ to the antichain $A$. These elements are incomparable to all elements in $A$ and incomparable to one another. Therefore, $A'$ is an antichain. It is easy to see that it is maximal in $X$. Furthermore, the intersection of $A'$ with the set $\{x \in [m]^n : x_1 = m\}$ is exactly $A$.
\end{proof}

Since $D_F(m,n)$ is nondecreasing with  $m$ and  $n$, we may conclude in particular that the number of maximal antichains in $X$ is at least the number of antichains in $[2]^{n-1}$, which is Dedekind number $D_E(2,n-1)$. Table 2 suggests that this bound is very weak. It also suggests that $D_F(m,n)$ grows extremely fast with $n$ even for $m=2$.

\subsubsection{Some numerical results}

Table~\ref{table.examples.size.mn} presents some values of $D_{F}(m,n)$ for small values of $m$ and $n$, computed with the help of the software system Macaulay2 \citep{M2}. For $[3]^{3}$, we used the function \texttt{maximalAntichains} provided by the package \texttt{Posets} in the software system Macaulay2 \citep{M2} and manually checked the result. For $[4]^{3}$ and $[3]^{4}$, we also used the function \texttt{maximalAntichains}, but without manual check.
For larger values (except when $m=2$ or $n=2$), the calculations are prohibitively long (indicated by question marks in Table~\ref{table.examples.size.mn}).

\begin{table}[hbp]
    \centering
    \begin{tabular}{r|rrrr}
        \hline
	$D_{F}(m,n)$ 	& $n=1$ 	& 2  	& 3 	  	& 4           \\
        \hline
        $m=1$		&	1 	& 1 	& 1	  	& 1            \\
        2			&	2 	& 3 	& 7	  	& 29           \\
        3			&	3 	& 9 	& \bf{144} 	& \bf{116547}   \\
        4			&	4 	& 27 	& \bf{10631}	&  ?           \\
        5			&	5 	& 83 	& 	? 	&   ?         \\
        \hline
    \end{tabular}
    \caption{Number of maximal antichains ($D_{F}(m,n)$) for small values of $m$ and $n$. Boldface entries are new.}
    \label{table.examples.size.mn}
\end{table}

For $[3]^{3}$, using \eqref{eq.BermanKohler}, we find $D_{E}(3,3)= 980$ so that the ratio $D_{F}(3,3)/D_{E}(3,3)$ is equal to $0.14693878$.
Similarly, for $[4]^{3}$,  we obtain $D_{E}(4,3)= 232848$ so that the ratio $D_{F}(4,3)/D_{E}(4,3)$ is equal to $0.04565639$, which implies a huge difference of expressivity between $F^{\underline{u}}$ and $E^{\underline{u}}$.

\section{Conclusion}
\label{se:conclusion}

Although our results about \nTBpd and its special cases are very partial, we have axiomatic and combinatorial results showing that
\begin{enumerate}

\item the analysis of the pseudo-disjunctive models is far more complex than that of the pseudo-conjunctive models;

\item there is a whole variety of pseudo-disjunctive models that are not all equivalent, contrary to what we observed for pseudo-conjunctive models;

\item most pseudo-disjunctive models are strict special cases of the corresponding pseudo-conjunctive models;

\item the pseudo-disjunctive model $F^{\underline{u}}$ is much more restrictive than the corresponding pseudo-conjunctive model.

\end{enumerate}
The strong asymmetry between the pseudo-conjunctive and pseudo-disjunctive models can be ascribed to the central role played by the relation $\rP$ in the definition of \nTBpd while $\rS$ is central in \nTBpc. Indeed, \citet{BouyssouPirlot2015AOR,BouyssouPirlot2015ITOR}
have shown that the nature of the relation $\rP$
is rather different from that of the relation $\rS$ in the ELECTRE methods.

 Hence, paralleling \citet{BouyssouMarchant2015}, we suggest to define  the \emph{dual} of \nTBpc not by means of \eqref{eq:MM}, but rather  by
\begin{equation}\label{eq:G}
x \in \Uu  \iff
\begin{cases}
p \rS x    &\text{ for some } p \in \Pro \quad \text{and}\\
\Not{x \rP q}  &\text{ for all } q \in \Pro.
\end{cases}
\end{equation}
It is easy to see that \nTBpc  and its dual
now correspond via the transposition operation consisting in inverting
the direction of preference on all criteria and permuting $\Aa$ and $\Uu$
\citep[see][]{AlmeidaDiasFigueiraRoy2010,BouyssouMarchant2015,Roy2002ETRI}.

Mimicking \citet[Th.\ 15]{BouyssouMarchantPirlot2023}, it is clear
this dual model is characterized by Linearity. Instead of taking
$\As$ as the set of profiles to delimit $\Aa$, we now take $\Us = \Max(\qodom, \Uu)$
to delimit the category
$\Uu$, still using ${\rSi} = {\relsi}$ and $\F = \{N\}$. 

If we replace \eqref{eq:MM} by \eqref{eq:G} in the definition of Models $F, F^{c}$ and $F^{u}$, it is also simple to see that they are all equivalent to the  dual  of \nTBpc.

\addcontentsline{toc}{section}{References}
\bibliographystyle{plainnat}
\bibliography{compens}
%
\end{document}